\documentclass[%
 reprint,
 aps,
 prl,
 floatfix,
]{revtex4-2}

\usepackage{graphicx}
\usepackage{dcolumn}
\usepackage{bm}
\usepackage{url}            
\usepackage{amsmath}
\usepackage{amssymb}
\usepackage{mathtools}
\usepackage{xfrac}
\usepackage{gensymb}

\begin{document}

\title{Maximizing MeV x-ray dose in relativistic laser--solid interactions}

\author{Kyle G. Miller}
\affiliation{Departments of Physics and Astronomy and of Electrical and Computer Engineering, University of California, Los Angeles, California 90095, USA}
\email{kylemiller@physics.ucla.edu}

\author{Dean R. Rusby}
\affiliation{Lawrence Livermore National Laboratory, Livermore, California 94551, USA}

\author{Andreas J. Kemp}
\affiliation{Lawrence Livermore National Laboratory, Livermore, California 94551, USA}

\author{Scott C. Wilks}
\affiliation{Lawrence Livermore National Laboratory, Livermore, California 94551, USA}

\author{Warren B. Mori}
\affiliation{Department of Physics and Astronomy, University of California, Los Angeles, California 90095, USA}

\date{\today}

\begin{abstract}
Bremsstrahlung x-rays generated in laser--solid interactions can be used as light sources for high-energy-density science.
We present electron and x-ray spectra from particle-in-cell and Monte Carlo simulations, varying laser pulse intensity and duration at fixed energy of 200$\,$J.  Superponderomotive electron temperatures are observed at low intensity; a new temperature scaling is given that depends on pulse duration and density scale length.  Short, high-intensity pulses create low-divergence electron beams before self-generated magnetic fields evolve, resulting in more forward-going MeV x-rays.
\end{abstract}

\maketitle

The interaction of a high-intensity laser pulse with a solid target produces a high forward flux of multi-MeV electrons through a variety of physical mechanisms~\cite{Wilks1992AbsorptionPulses,Malka1996ExperimentalTarget,Wilks1997AbsorptionPlasmas,Pukhov1998RelativisticSimulations,Santala2000EffectInteractions,May2011MechanismInterface}.  The ``hot'' electrons can produce bremsstrahlung x-rays~\cite{Perry1999Hardinvited,Courtois2009EffectPulses}, radiation that has been considered for a wide variety of applications~\cite{Kieffer2002FutureImaging,Rusby2016PulsedDriver,Brenner2016Laser-drivenAccelerators}.  Work has been devoted to optimizing the bremsstrahlung production and controlling its directionality, including the use of pre-formed plasmas~\cite{Santala2000EffectInteractions,Courtois2009EffectPulses,Culfa2016PlasmaPlasmas}, advanced nanowires~\cite{Jiang2014EnhancingStructures,Ebert2020} and cone-shaped targets~\cite{Gaillard2011IncreasedTargets,Kluge2012HighMechanisms}.  In addition, laser absorption has been shown to rise with increased laser intensity and oblique incidence~\cite{Ping2008AbsorptionRegime,Davies2009LaserRegime}, and particle energies greater than predicted by traditional scaling laws have been observed for long-pulse lasers~\cite{Simpson2021ScalingRegime}.

In this letter we present particle-in-cell (PIC) simulation results on relativistic laser--solid interactions where the laser pulse amplitude and duration are varied while holding energy constant.  Several recent computational techniques~\cite{May2014,Miller2021a,Fonseca2013} are employed that allow for large-scale (hundreds of $\mu$m), long-time (tens of ps) simulations of such interactions.  Electron spectra and beam profiles determined from PIC simulations are used as input for Monte Carlo simulations to calculate x-ray spectra, and a clear maximum in 1--5 MeV x-ray dose is found as a function of laser intensity.

\begin{figure}
    \centering

     \includegraphics[width=\linewidth]{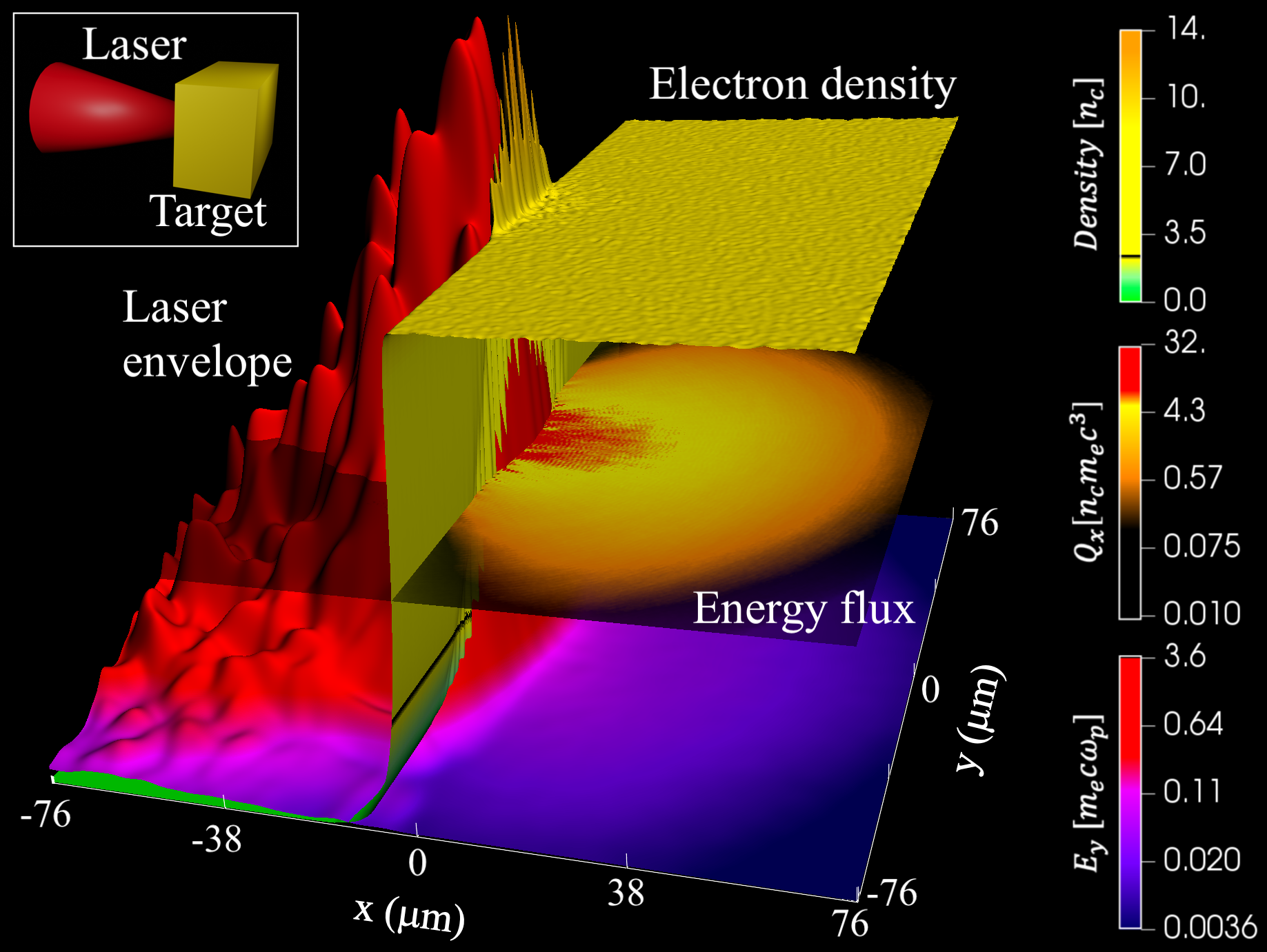}
    \caption{Electron density, transverse laser field envelope and forward electron energy flux for a laser with $(a_0,\tau)=(1.83,3\,\mathrm{ps})$ and $w_0$ of 30$\,\mu$m near when the peak of the laser pulse reaches the critical surface.  Quantities are spatially averaged for visualization.}
    \label{fig:2d-sim}
\end{figure}

All PIC simulations are performed with \textsc{Osiris}~\cite{Fonseca2002}.  We use an exponential density ramp of the form $10\,n_c e^{-x/L_0}$ over the region $x \in [-27.6,0]\,\mu$m, where the scale length, $L_0$, is set to 3$\,\mu$m and $n_c$ is the critical density.  The density is set to zero and $10\,n_c$ on the left and right of this region, respectively; the critical density is thus located at $-6.9\,\mu$m. The laser is launched from the left boundary in the $\hat{x}$-direction.
We use two-dimensional simulations where the plasma is $240\,\mu$m wide in $y$ and that extends to 150$\,\mu$m in $x$ at the right edge, with open and thermal boundary conditions (BCs) on all sides for fields and particles, respectively (except for absorbing particle BC at left wall in $x$).  Although each PIC simulation corresponds to a family of cases where the normalized parameters are the same, we give physical units that correspond to the incoming laser having a wavelength of 1$\,\mu$m. A diffraction-limited Gaussian laser with spot size, $w_0$, of 30$\,\mu$m is launched from left to right with a focus at the critical surface (the Rayleigh length is $\pi w_0^2/\lambda$).
We vary the amplitude and duration of the laser, keeping the product $a_0^2\tau$ constant at 10$\,$ps, where $a_0 \simeq 8.6\times 10^{-10} \sqrt{I_0\,[\mathrm{W}/\mathrm{cm}^2]} \lambda\,[\mu\mathrm{m}]$ is the normalized vector potential for vacuum intensity $I_0$ and $\tau$ is the full width at half maximum (FWHM) of intensity.  The lowest-intensity case has $(a_0,\tau)=(0.58,30\,\mathrm{ps})$, while the highest-intensity has $(a_0,\tau)=(31.6,0.01\,\mathrm{ps})$, corresponding to an energy of $200\,$J for all cases; this energy is achievable by many picosecond-class laser systems.  The laser profile is Gaussian in the transverse direction and uses a polynomial fit to a Gaussian temporally.  Cell sizes are $0.4\,c/\omega_0$ in each direction, where $\omega_0$ is the laser frequency ($c/\omega_0 = 159.2\,$nm).  (A limited set of simulations are also performed with smaller cell sizes $0.2\,c/\omega_0$ in each direction and with a peak density of $30\,n_c$ to ensure that the results are consistent for smaller cell sizes and not subject to relativistic transparency.)   The time step is $0.282\,\omega_0^{-1}$, and simulations are run to 2$\,$ps after the laser is extinguished.  Figure~\ref{fig:2d-sim} shows the electron density, laser envelope and forward electron energy flux (defined as $Q_x = \int v_x(\gamma-1)m_e c^2 F(\vec{x},\vec{p}) \,d\vec{p}$ for the plasma distribution function $F$) for a simulation with $(a_0,\tau)=(1.83,3\,\mathrm{ps})$ at time 0.5$\,$ps after the laser peak.  The plasma density sharply increases near the laser axis, and the laser is observed to penetrate slightly into the overdense region where the electron energy flux is greatest.

To make such long-time simulations of relativistic laser--solid interactions possible, we employ several unique PIC techniques.  Electrons (ions) are simulated with 64 (16) particles per cell, but to avoid numerical stopping of energetic particles from the enhanced wakes from macroparticles~\cite{May2014}, fast electrons ($\gamma > 1.5$) are split into 64 smaller particles.  To prevent hot-electron refluxing from simulation boundaries, extended particle absorbers~\cite{Miller2021a} are employed over the regions $|y|\geq 110\,\mu$m and $x\geq 130\,\mu$m to gradually stop energetic particles without large electric field growth.  Low-energy particles with small charge in the absorbing regions are combined with other particles in the same octant of momentum space to reduce computational load.  The left vacuum boundary in the $\hat{x}$-direction is placed far from the vacuum-plasma interface so as to be causally separated (varied based on simulation time), and dynamic load balancing~\cite{Fonseca2013} is performed every 50~time steps. Last, to avoid numerical grid heating and to reduce the enhanced wakes, we use cubic particle shapes.

\begin{figure}
    \centering

     \includegraphics[width=\linewidth]{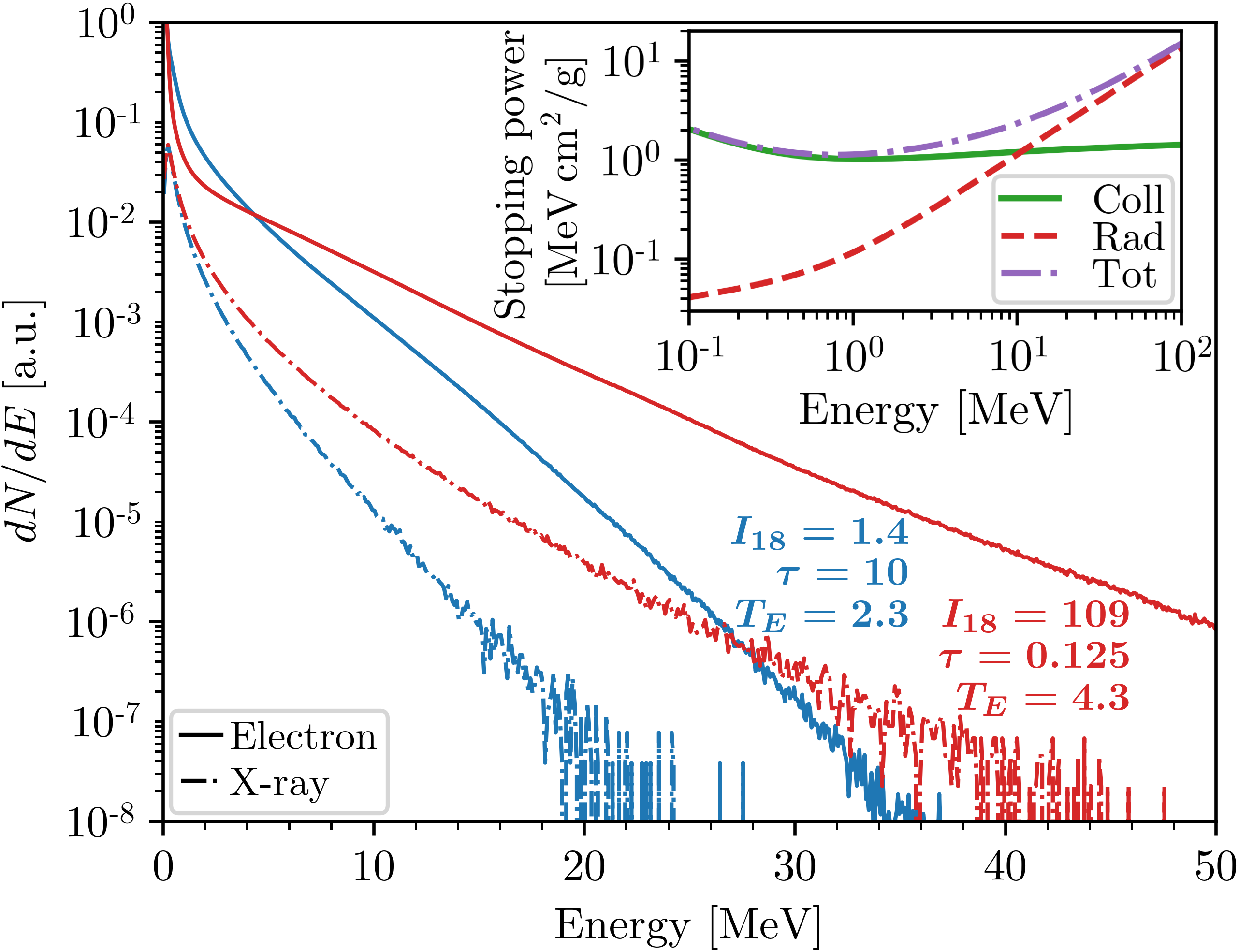}
    \caption{\label{fig:spectra} Time-integrated electron and resulting photon spectra for two simulations at identical laser pulse energy (laser intensity, FWHM and electron temperature labeled in units of $10^{18}\,$W/cm$^2$, ps and MeV, respectively). Inset shows the stopping power (collisional, radiative and total) of tungsten as a function of electron energy.
    }
\end{figure}

Assuming high-energy electrons deep inside the plasma will continue on through the target, the Monte Carlo simulation package \textsc{Geant}4~\cite{Agostinelli2003Geant4aToolkit} is used to compute the energy and number of bremsstrahlung x-rays generated by these electrons.  The \textsc{Geant}4 code is capable of simulating the bremsstrahlung radiation from individual particles, though it does not include self-consistent electromagnetic fields or other collective effects.  The forward-going electron spectra are first extracted from \textsc{Osiris} every 200 time steps over the region 50--57.5$\,\mu$m in $x$.  Next, mono-energetic electrons between the energies of 0.2--100$\,$MeV are injected into 500$\,\mu$m of tungsten, and the emitted x-ray spectra are gathered as a function of emitted angle.  Time-integrated, forward-going PIC electron spectra are then interpolated onto the mono-energetic electron data and weighted to find the corresponding x-ray spectra.

Figure~\ref{fig:spectra} shows the forward-going \textsc{Osiris} electron spectra from the $x$ range given previously for two simulations.
The inset shows the stopping power (collisional, radiative and total) of tungsten as a function of mono-energetic electron energy.  Dash-dotted lines show the resulting x-ray spectra computed from \textsc{Geant}4 and collected over all angles.  Although the low-intensity (blue) simulation in Fig.~\ref{fig:spectra} gives nearly twice the number of 1--5$\,$MeV electrons as the high-intensity (orange) simulation, the large stopping power of many-MeV electrons results in more x-rays of all energies for the high-intensity simulation with a higher electron temperature.

\begin{figure}
    \centering

     \includegraphics[width=\linewidth]{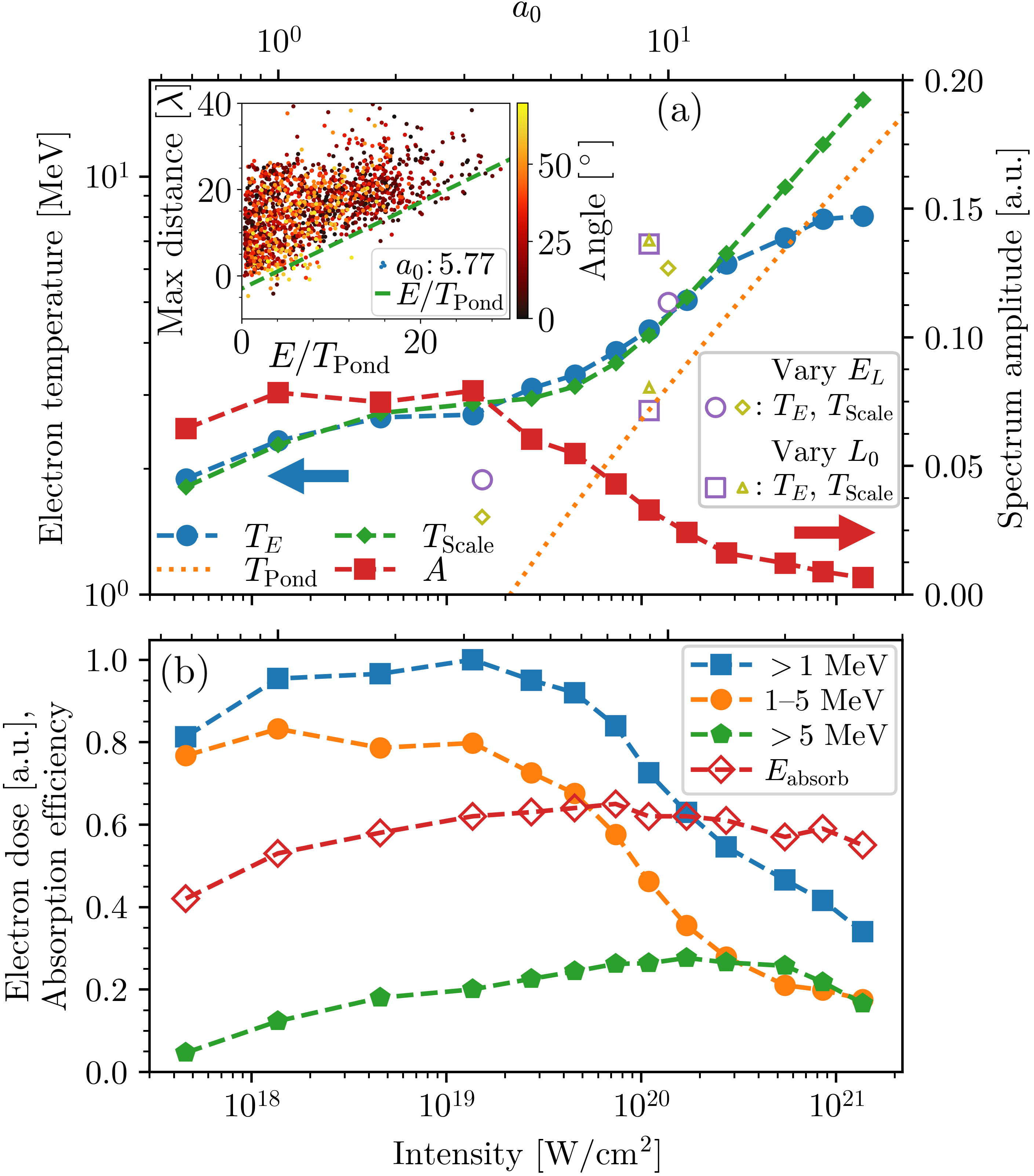}
    \caption{\label{fig:temps} (a)~Scaling of hot electron temperature with laser intensity at constant energy.  The $A$ and $T_E$ values result from an exponential fit $Ae^{-E/T_E}$ to the simulated cumulative forward electron spectra, where $E$ is energy. Inset shows the maximum distance tracked particles extend away from the constant-density region as a function of final energy for a single simulation, colored by absolute angle from the $x$-axis.  (b)~Electron dose (normalized to largest value) and fraction of the laser energy absorbed ($E_\mathrm{absorb}$) for all simulations.
    }
\end{figure}

Aggregate data from our suite of PIC simulations is shown in Fig.~\ref{fig:temps}.  In (a) we show the amplitude, $A$, and electron temperature, $T_E$, of the exponential fits done over 3--30$\,$MeV for each run.  The generation of energetic electrons is a complicated process that has been described by many differing scaling laws and analyses~\cite{CompantLaFontaine2014}.
Rather than compare to this body of work, as a point of reference we also show the ponderomotive temperature scaling~\cite{Wilks1993SimulationsInteractions,Malka1996ExperimentalTarget}, which goes as $T_\mathrm{Pond} = m_ec^2 (\sqrt{1+a_0^2/2}-1)$ for a linearly polarized laser. 
The ponderomotive scaling indicates that $T_E$ scales with the laser field for $a_0\gg1$. Such an  energy corresponds to the work done on an  electron after it moves a single laser wavelength, which is more characteristic of the high-intensity laser pulses with short duration.
However, at lower intensity (long duration), the plasma expands and electrons in the low-density region can repeatedly interact with the laser over a large distance, generating electrons with temperatures and energies much larger than the ponderomotive potential~\cite{Mendonca1982StochasticityWaves,Forslund1985Two-DimensionalHeating,Meyer-ter-Vehn1999,Sorokovikova2016GenerationPlasmas,Kemp2020}.

Insight into how the spectra of energetic electrons are generated and how they scale with intensity can be found by looking at the tracks of individual electrons. The inset to Fig.~\ref{fig:temps}(a) shows the maximum distance a sampling of 1500 tracked electrons travels into the low-density region as a function of final energy for the simulation with $(a_0,\tau)=(5.77,300\,\mathrm{fs})$.  On this scale the initial density ramp begins at 27.6$\,\lambda$ and becomes constant at 0$\,\lambda$, though the constant-density shelf is pushed back to $-8\,\lambda$ near the laser axis late in the simulation.
For a given energy, there is a distribution of distances that electrons have penetrated into the low-density region, leading to an electron temperature, $T_E$. However, to obtain an energy of at least $NT_\mathrm{Pond}$, an electron must interact with the laser for a distance of $N\lambda$. This is seen by the dashed line with a slop of unity.  Electrons are colored by the angle of their trajectory with respect to the $x$-axis.  Though no trend in angular distribution is observed with distance traveled into the pre-plasma, we see that more energetic electrons have on average a lower divergence. We discuss this shortly.

To better predict electron temperature based on the laser and plasma conditions, we look for a temperature model that depends on both the laser duration, $\tau$, and initial scale length of the pre-plasma, $L_0$.
As laser duration and plasma scale length decrease to zero, the model should approach the ponderomotive temperature, since this behavior has been seen in a variety of work~\cite{Wilks1992AbsorptionPulses,Lasinski1999Particle-in-cellApplications,Tonge2009ALasers}.  We thus propose a scale temperature
\begin{equation}
T_\mathrm{Scale} \propto \left[\sqrt{1+\left( \tau/\Bar{\tau} \right)^2} \right]^p,
\end{equation}
where the overbar indicates some reference duration and $p$ is an exponent.  With this dependence on $\tau$, the expression equals unity and has zero derivative at $\tau=0$; we likewise propose the same proportionality for the scale length.

The proposed temperature scaling is then of the form
\begin{equation} \label{eq:tscale}
T_\mathrm{Scale} = T_\mathrm{Pond} \left[1+\left( \tau/\Bar{\tau} \right)^2 \right]^{\sfrac{p}{2}} \left[1+\left( L_0/\Bar{L}_0 \right)^2 \right]^{\sfrac{q}{2}}.
\end{equation}
Performing a fit of the simulation data to this model yields $(\Bar{\tau}, p, \Bar{L}_0, q) = (250\,\mathrm{fs}, 0.72, 0.73\,\mu\mathrm{m}, 0.23)$.  The fit is plotted as $T_\mathrm{Scale}$ in Fig.~\ref{fig:temps}(a) and is shown to accurately represent the data for all but the highest intensities.  The three simulations with highest intensity have lasers with $\tau \leq 25\,$fs, or fewer than 16 full laser cycles, meaning that many hot electrons are generated when the laser is far from peak intensity.
Additional simulations are performed with varied laser energy, $E_L$, and initial density scale length, $L_0$, as pictured by the hollow symbols in Fig.~\ref{fig:temps}(a).
Although $T_\mathrm{Scale}$ predicts the observed electron temperatures with reasonable accuracy for these differing cases, some discrepancy illustrates that the $(\Bar{\tau}, p, \Bar{L}_0, q)$ coefficients may vary for drastically differing plasma density and laser spot size, wavelength and energy.

In Fig.~\ref{fig:temps}(b) we report the time-integrated, forward-going electron dose for three energy ranges (normalized to the largest value shown), along with the fraction of the laser energy absorbed by all forward-going electrons.  We show the electron dose in these energy ranges since we desire a large number of x-rays in the 1--5$\,$MeV range.  The dose of 1--5$\,$MeV electrons is largest for the low-intensity laser pulses; and with increasing laser energy the number of electrons above 5$\,$MeV steadily increases, while the total number of MeV electrons decreases.  However, we observe that the total energy absorbed into the hot electrons from the laser remains close to 60\% for all but the lowest-intensity cases.  The near-constant absorption poses the question of how to best maximize x-ray production given the observed electron doses.

\begin{figure}
    \centering

     \includegraphics[width=\linewidth]{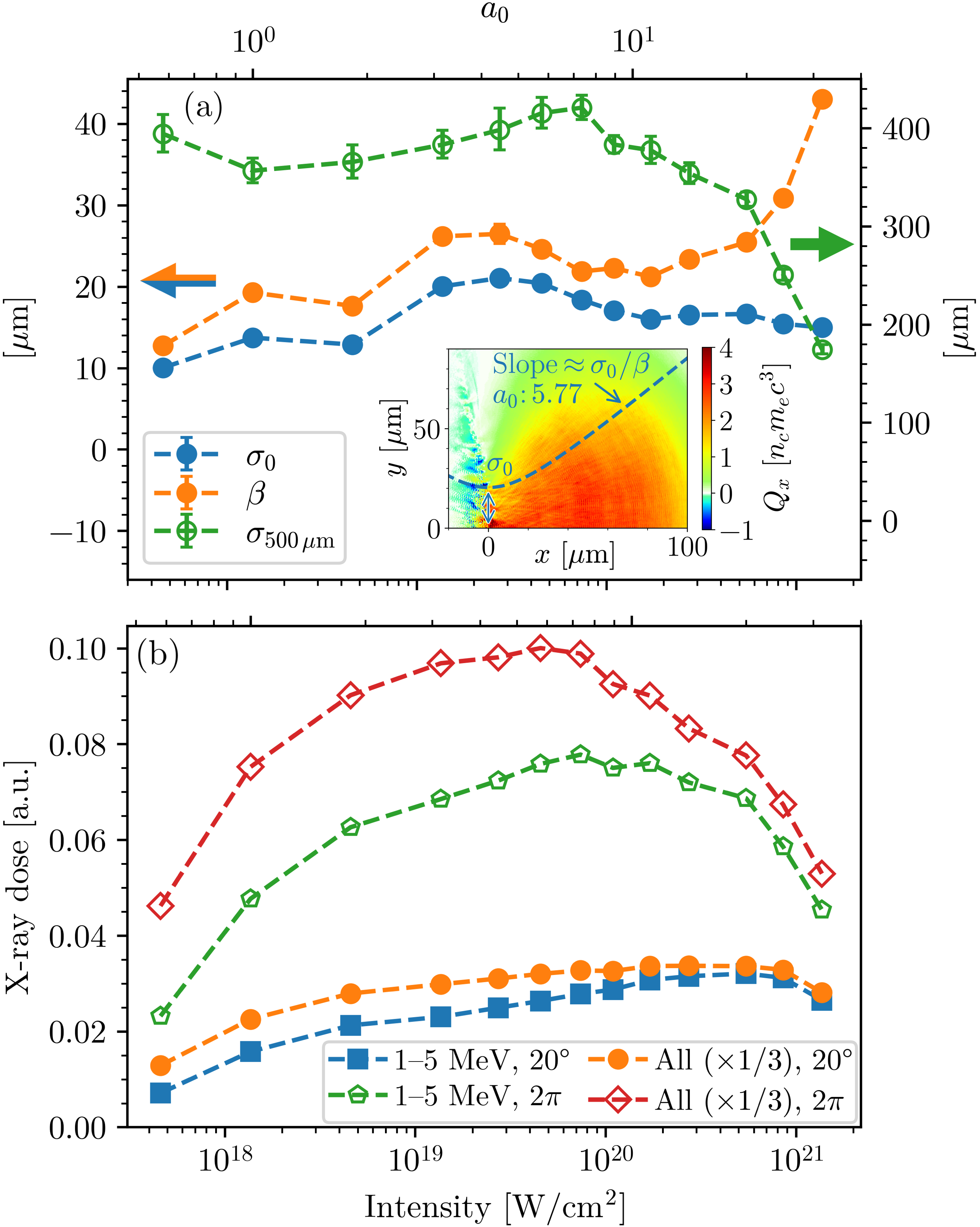}
    \caption{\label{fig:x-ray} (a)~Beam parameters and spot size at 500$\,\mu$m for the hot electrons with respect to laser intensity, where $\sigma(x) = \sigma_0 \sqrt{1+(x/\beta)^2}$.  Inset shows example fitted beam parameters.
    (b)~X-ray dose as a function of intensity with the same normalization as in Fig.~\ref{fig:temps}(b), collected within a forward cone of 20$\degree$ aperture (blue/orange) and at all angles (green/red).  X-ray counts for all energies are scaled by $1/3$ for visibility.
    }
\end{figure}

To accurately determine the x-ray production using such a large variety of pulse amplitudes, we must first characterize any differences in the angular spread of the electron beam.  To accomplish this we track a 10-$\mu$m slice of electrons generated from the peak of each laser pulse as it travels through and spreads into the target from 0 to 130$\,\mu$m.  Fitting the forward energy flux near the vacuum interface to a Gaussian profile in the transverse direction ($e^{-y^2/2\sigma_0^2}$), we compute the initial beam width, $\sigma_0$.  Assuming a beam evolution of $\sigma(x)=\sigma_0 \sqrt{1+(x/\beta)^2}$, we then make a fit for $\beta$ using the energy flux as a function of $x$. Although such a fit is most useful when the incoming energy is in the form of a well-defined beam with low energy spread and emittance, it still provides a useful characterization for the beam.  We show these parameters for the various simulations in Fig.~\ref{fig:x-ray}(a), along with the spot size evaluated at 500$\,\mu$m.  The electron beam becomes significantly less divergent for the high-intensity, short-duration laser pulses, which is favorable for forward x-ray emission.

From the simulated electron spectra and beam profiles, we then compute the observed x-ray spectra using \textsc{Geant}4 as described previously.  In Fig.~\ref{fig:x-ray}(b), we plot the x-ray dose collected both inside a forward cone of 20$\degree$ aperture as well as over an entire spherical surface.  The energy range of 1--5$\,$MeV is shown alongside all energies, where the latter has been scaled by $1/3$ for visibility.  For context, computing the x-ray spectra using only the amplitude and temperature fits shown in Fig.~\ref{fig:temps}(a) reproduces the 1--5$\,$MeV x-ray dose to 1\% accuracy (5\% accuracy for the highest- and lowest-intensity cases).  The forward-going x-ray dose very closely follows the $>5\,$MeV electron dose shown in Fig.~\ref{fig:temps}(b), peaking somewhere between $10^{20}$--$10^{21}\,$W/cm$^2$.  The 1--5$\,$MeV x-ray dose peaks near $5\times10^{20}\,$W/cm$^2$, demonstrating that high-energy ($>5\,$MeV) and low-divergence electrons are desirable for producing few-MeV x-rays.

\begin{figure}
    \centering

     \includegraphics[width=\linewidth]{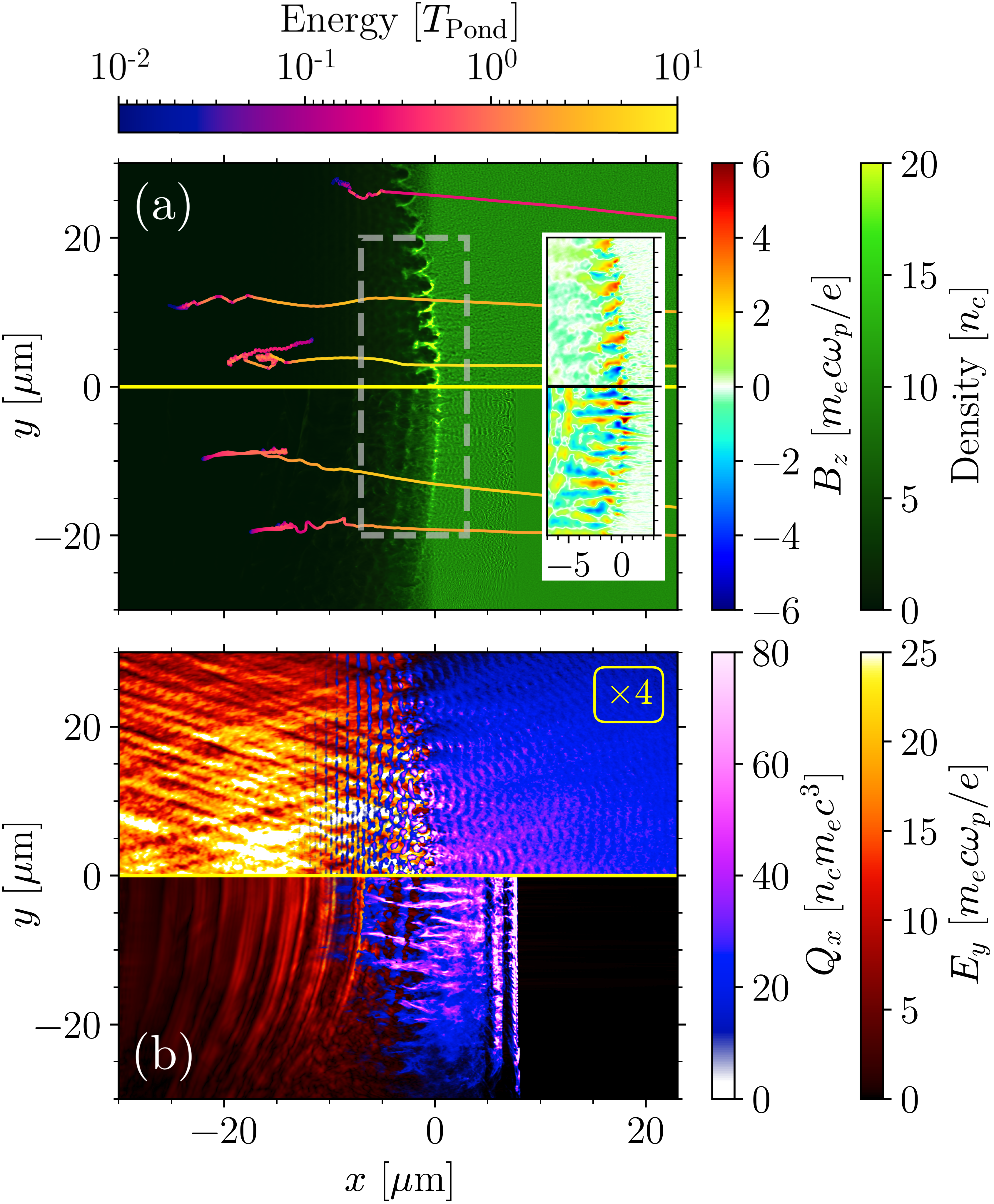}
    \caption{\label{fig:fld-tracks} (a)~Electron density and time-averaged magnetic field (inset for the indicated region) for two simulations, along with the tracks of several electrons. (b)~Transverse electric field envelope and forward electron energy flux for the same simulations, with the top pane scaled by four for visualization.  In each split-pane figure, the top and bottom halves show the $(a_0,\tau)=(5.77,300\,\mathrm{fs})$ and $(25,16\,\mathrm{fs})$ cases, respectively.
    }
\end{figure}

To investigate the decreasing electron divergence at high intensity, we show in Fig.~\ref{fig:fld-tracks}(a) and (b) electron density, time-averaged magnetic field, electric field envelope and forward electron energy flux for long- (top) and short- (bottom) duration lasers at different times.  Note that the far-field electron spot size for the long-duration laser is nearly double that of the short-duration laser.  Both panels in (a) show large self-generated surface magnetic fields from a Weibel-like instability~\cite{Fiuza2012Weibel-instability-mediatedLasers,Huang2019MaximizingInteractions,Shukla2020InterplayInteractions}, but the density perturbation has progressed more in the top pane.  However, in the bottom pane of (b), the first two bunches of electrons (spaced approximately two laser wavelengths apart) are observed to propagate through the plasma with minimal divergence.  The maximum cold-fluid-limit growth rate of the Weibel-like instability dominated by oblique modes~\cite{Bret2010ExactFunctions} is given as $\Gamma = \sfrac{\sqrt{3}}{2^{4/3}}\left(\sfrac{n_b}{\gamma_b n_p}\right)^{1/3}\omega_p$, where $n_b$ and $\gamma_b$ are the density and relativistic factors of the beam, and $n_p$ and $\omega_p$ correspond to the bulk plasma.  The simulation for the bottom panels of Fig.~\ref{fig:fld-tracks} has $(n_b/n_p,\gamma_b) = (6\times10^{-4},12)$, yielding a growth rate of $\Gamma^{-1} \approx 7\,$fs, or about two laser periods.
Tracks of individual particles chosen based on their energy deep in the plasma are shown in (a), where we observe that electrons in the top pane are deflected by surface magnetic fields before free streaming through the remainder of the plasma.
Overall, the shortest laser pulses exhibit decreased electron divergence since many energetic electrons are generated prior to the formation of self-generated magnetic fields.

In conclusion, we have studied the generation of MeV x-rays from intense laser--solid interactions using a combination of large-scale, two-dimensional, high-fidelity \textsc{Osiris} PIC and \textsc{Geant}4 Monte Carlo simulations.  Keeping the laser pulse energy constant at 200$\,$J for a 30-$\mu$m spot size, we find that the optimal forward dose of 1--5$\,$MeV x-rays results from a laser intensity near $5\times10^{20}\,$W/cm$^2$ and duration 25$\,$fs.
A new scaling of the energetic electron temperature given in Eq.~(\ref{eq:tscale}) includes a dependence on laser duration and initial density scale length.
The formation of Weibel-like magnetic fields near the critical-density surface over only a few laser periods increases the electron beam divergence, leading to decreased forward-going x-ray emission.
Based on simulation results for a wide range of laser pulse parameters, we find that, for fixed laser energy, high-intensity (and short-duration) laser pulses generally produce more few-MeV x-rays than do low-intensity pulses due to the formation of electron beams with high temperature and low divergence.

\begin{acknowledgments}
This work was performed under the auspices of the U.S. DOE by Lawrence Livermore National Laboratory (LLNL) under Contract DE-AC52-07NA27344 and funded by the LLNL LDRD program with tracking code 19-SI-002 under Contract B635445. Additional support was given by DOE grants DE-SC0019010 and DE-NA0003842 and NSF grant 2108970.  Release under the IM:LLNL-JRNL-831656-DRAFT.  Computer simulations were performed with a Grand Challenge allocation at the Livermore Computing Center and on NERSC's Cori cluster (account m1157).
\end{acknowledgments}

\bibliography{references}

\end{document}